\documentclass[preprint,aps,showpacs,floats]{revtex4}
\usepackage{graphicx}
\usepackage{amsmath}
\newcommand{\be}{\begin{equation}}              
\newcommand{\ee}[1]{\label{#1} \end{equation}}  
\newcommand {\eps} {\epsilon}
\newcommand{\bee}{\begin{eqnarray}}             
\newcommand{\eee}{\end{eqnarray}}               

\usepackage{color}
\begin{document}

\title{Flatons: flat-top solitons in extended Gardner equations}

\author{Philip~Rosenau}
\affiliation{School of Mathematics,\\
 Tel-Aviv University, Tel-Aviv 69978, Israel}
\email[]{rosenau@tauex.tau.ac.il}

\author{Alexander~Oron}
\affiliation {Faculty of Mechanical Engineering,\\
Technion - Israel Institute of Technology, Haifa 3200003, Israel}
\email[]{meroron@technion.ac.il}

\date{\today}

\begin{abstract}

In both the Gardner equation and its extensions, G(n):
$u_{t} +(u^{2}-u^{3})_{x}+(u^{n})_{xxx}=0~$, $n \ge 1$,
the non-convex convection bounds the range of solitons/compactons
velocities beyond which they dissolve and kink/anti-kink form.
Close to solitons barrier we unfold a narrow strip of
velocities where solitons shape undergoes a structural change and
rather than grow with velocity, their top flattens and they widen
rapidly;  $\epsilon^{2} \ll 1$
change in velocity causes their width to expand $\sim \ln{1/\epsilon}$.
To a very good
approximation these solitary waves, referred to as flatons, may be viewed as made of a kink and anti-kink 
placed at an arbitrary distance from each other.
Like ordinary solitons, once flatons form they are
very robust. A multi-dimensional extension
 of the Gardner equation reveals that spherical flatons are as prevalent and in many cases
 every admissible velocity supports an entire
 sequence of multi-nodal flatons.
\end{abstract}
\maketitle

\section*{1. Introduction}

This communication is concerned with both one-dimensional and
multi-dimensional flatons which are solitary
formations with an almost flat top that emerge in
dispersive systems wherein the prevailing mechanisms
bound the range of velocities at which solitons are admissible. Beyond the barrier
 solitons dissolve into a constant and kink and anti-kink emerge.
It will be seen that close to the barrier there is a transition layer where solitons undergo a structural change and rather
than to grow with velocity, they widen very quickly
and, as their velocity closes on its upper bound, they become
almost completely flat at their top. It is thus natural to refer to such solitons as flatons: velocity changes
$~{\cal{O}}(\eps^{2})$ cause flatons to
expand as $\sim ~ - \ln{\eps}$). To a very good approximation
1D flatons may be viewed as made of a kink and anti-kink pair
which travels at the transition velocity of their range,
 placed at an arbitrarily chosen distance from each other.

Among the many dispersive systems which may support flatons, perhaps
the simplest one is given via an extended Gardner equation G(n)
\begin{equation}
 G(n):~~~~u_{t} + (u^{2}-u^{3})_{x}+(u^{n})_{xxx}=0,~~~  n \ge 1, \label{23n}
\end{equation}
\noindent
which for $n=1$ reduces to the classical Gardner equation \cite{Drazin}, \cite{Gardner},
whereas for $n>1$ it becomes a
non-convex extension of the $K(m=2,3;n)$ equation which yields compactons.
\cite{RosHym,RZ}.
Though the  Gardner equation was originally introduced as an auxiliary
mathematical device \cite{Gardner} used to deduce an infinity of
conservation laws,
it has since emerged in a variety of applications. For instance, the
convection's non-convexity may be due to two opposing
mechanisms causing the convection to reverse its direction at a
critical amplitude, as may happen
during liquid film's flow on a tilted plane when
gravity and Marangoni stress induce convection of opposing
directions \cite{Wit}. Thus in Eq. (\ref{23n}), convection's velocity
$C(u)=2u-3u^{2}$ assumes its maximal value at $u=1/3$ and
reverses its direction at $u=2/3$.

As a further extension, we shall consider other
nonlinearities and higher dimensions as well
\begin{equation}
G_{d}^{n}(\kappa,\ell):~~~~u_{t} + (c_{+}u^{\kappa}-c_{-}u^{\ell})_{x}+\big(\nabla^{2}u^{n}\big)_{x}=0, \label{ND}
\end{equation}
where $d=1,2,3$ stands for the spatial
dimension and $c_{\pm}, \kappa$ and $\ell$
are positive constants, and show that the extended equation
supports a far richer variety of flatons
than on the line. Actually, in certain cases to be detailed in Section III,
in particular when the underlying potential is symmetric,
every admissible velocity may support an
entire sequence of multi-nodal flatons. We note in passing that
within the complex Klein-Gordon realm spherical flat solitary formations
were addressed, among others, in \cite{Q-balls} and \cite{Q-balls2}.

The plan of the paper { is as follows}: in Section II,
we explore the emergence of
1D flatons and study their dynamics in the framework of the G(n) equation.
In Section III, we extend our study to the spherically
symmetric $G_{d}^{1}(\kappa,\ell)$ flatons.
Section IV concludes with discussion and comments.

\section*{2. Emergence of 1D flatons}

We start with the original 1D Gardner equation \cite{Gardner},
\begin{equation}
 G(1):~~~~u_t + \left(u^2-u^3\right)_x+u_{xxx}=0. \label{G1}
\end{equation}
Two integrations in the traveling frame with velocity $\lambda$, where $s=x-\lambda t$, yield
\begin{equation}
\frac12 u_s^2 + P_1(u)=0~~~{\rm with}~~P_1(u)=-\frac12 \lambda u^2 +
\frac13 u^3 - \frac14 u^4.  \label{G2}
\end{equation}

\begin{figure}
\label{poten}
\noindent \begin{centering}
\vspace*{4mm}
\includegraphics[width=20pc]{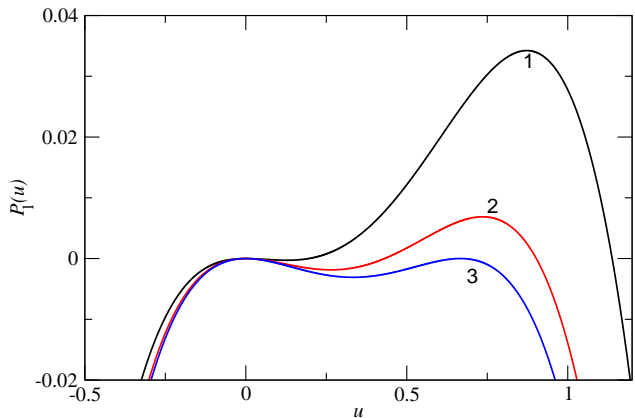}
\par\end{centering}
\vspace* {-3mm}
\caption{
Display of the potential
$P_1(u)=-\frac{1}{2}\lambda u^{2}+\frac{1}{3}u^{3}-\frac{1}{4}u^{4}$ for
Eq. (\ref{G1}) with $\lambda=1/9$, $7/36$ and $2/9$, cases 1-3, respectively.
Flatons emerge in the domain between cases 2 and 3.
In case 3 wherein $P=P'=0$, there are no solitons. { Instead
kink and/or anti-kink emerge.}
}
\end{figure}

Equation (\ref{G2}) yields solitons
\begin{equation}
u=\frac{3\lambda}{1+ \sqrt{1-\frac{9\lambda}{2}}
\cosh{\left(\sqrt{\lambda} (x- \lambda t)\right)}}
\label{TW2modkdv}
\end{equation}
with their peak reaching
\begin{equation}
u_{max}=\frac{2}{3}\Big(1- \sqrt{1-\frac{9\lambda}{2}} \Big).
\label{umax}
\end{equation}

The defocusing effect due to the cubic term enforces
 an upper bound, $\lambda =2/9$, on the admissible propagation
velocities. At the limiting velocity,
the soliton (\ref{TW2modkdv}) flattens into a constant.
This is a singular limit for, in addition, also kink and anti-kink emerge,
\begin{equation}
u=\frac{2}{3\left(1+ \exp( \mp \frac{\sqrt2}{3} s )\right)}~~~
{\rm{where}}~~~ s=x-\frac29 t.
\label{Kmodkdv}
\end{equation}

\begin{figure}
\noindent \begin{centering}
\includegraphics[width=20pc]{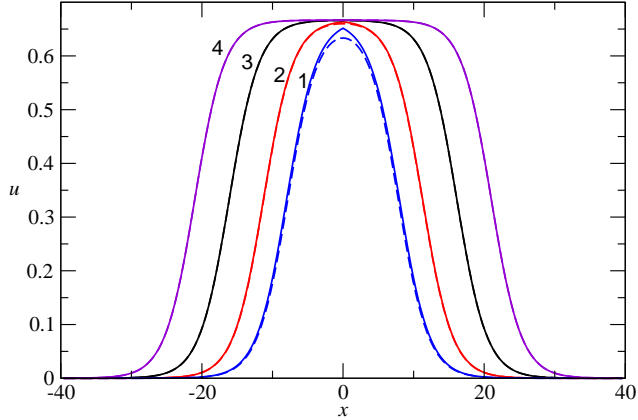}
\par\end{centering}
\caption{
Comparison between the
approximate shape (solid lines) of the G(1) flaton, Eq. (\ref{appflaton}),
and the exact solution (dashed lines) for a several values of
{$x_0$ (or $\epsilon$, see Eq. (\ref{x0eps}))}:
1- $x_0=8.0 $ ($\Rightarrow \epsilon =0.05$),
2- $x_0=11.27$ ($\Rightarrow  \epsilon =10^{-2}$),
3- $x_0=16.13$ ($\Rightarrow \epsilon =10^{-3}$),
4- $x_0=21.01$ ($\Rightarrow \epsilon =10^{-4}$).
Notably, apart of the first pair with visible difference at the top
between the exact solution and its approximation,
in other cases, they are truly indistinguishable.
}
\label{eq6eq8}
\end{figure}

The disconnection between solitons and kinks can be also seen
from the underlying potential $P_{1}$ in Fig. 1;
as $\lambda \rightarrow 2/9$, potential's top comes down, but once it
touches the u-axis, solitons dissolve and instead a kink and anti-kink emerge.

To unfold the 1D flatons we proceed as follows: first,
we join the kink with an anti-kink into a pair, with their centers
(defined as the location where $u$ assumes half of its maximal value)
placed at
$\pm x_{0}$, $x_{0} \gg 1$.
Clearly, for a finite $x_{0}$ such structure
\begin{equation}
u=\frac{2}{3\left(1 + \exp{\left[\frac{\sqrt2}{3}
\left(|x|-x_{0}\right)\right]}\right)} ~~~ -\infty<x<\infty.
\label{appflaton}
\end{equation}
is an approximate solution. It will be used as
an initial input for Eq. (\ref{G1}).

\begin{figure}
\label{G1in}
\noindent \begin{centering}
\hspace*{-4cm}
\vspace*{-4cm}
\includegraphics[width=55pc]{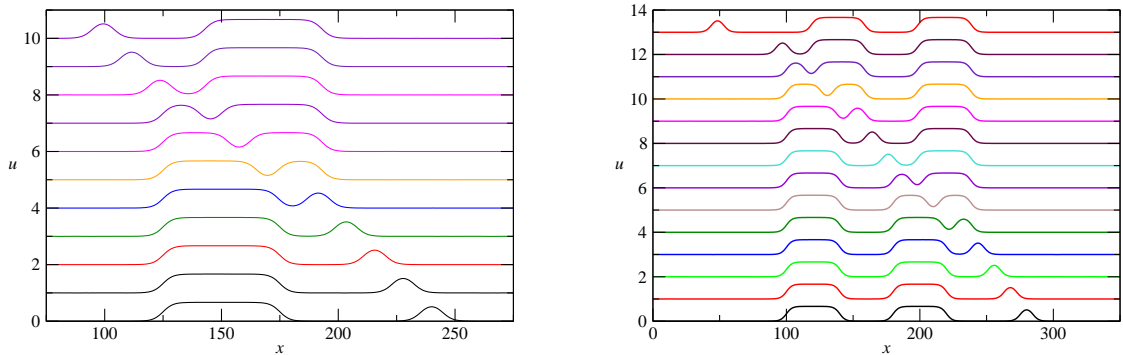}
\par\end{centering}
\vspace* {-21cm}
\caption{
Interaction between G(1) flatons and a soliton.
Left panel: Space-time display of the interaction of a wide
flaton, $x_0=25$, with a  $\lambda=0.21$ soliton,
shown here and elsewhere in a reference frame moving with the kink velocity of the flaton.
From the bottom to the top; the
time ranges from $t=0$ to $t=10^{4}$ with time intervals of  $1000$.
Right panel:
Space-time display of  the interaction between two
equal flatons, $x_0=20$, with a $\lambda=0.21$ soliton.
From the bottom to the top, the time ranges from $t=0$
to $12000$ with time intervals of $1000$ and finally $t=16000$.
}
\end{figure}

Given by Eq. (\ref{appflaton}) $u$ is continuous at $x=0$,
where $\displaystyle u \simeq 1- \exp{\left(-\frac{\sqrt2}{3} x_{0}\right)}$,
and though its derivative undergoes a finite jump
$[u] \simeq 2\exp{\left(-\frac{\sqrt2}{3} x_{0}\right)}$,
for $1<<x_{0}$ the jump is very small and
decreases very quickly with $x_{0}$. Indeed, as clearly seen in Fig. 2,
the approximate solution converges very quickly with $x_{0}$ to
the exact solution.

%
The solutions of the Gardner equation and other partial
differential equations considered, {\color{red} are} numerically determined
using periodic boundary conditions in a sufficiently large domain
employing the Newton-Kantorovich
procedure \cite{Boyd} which has second-order
accuracy in both time and space and an implementation similar to the
one in \cite{Dji1995}, for details see \cite{RO}.

To test the viability and robustness of flatons
we let them collide with solitons.
As seen in Fig. 3, they reemerge in their original form without any
observable debris,
with the left panel displaying the interaction between
a wide flaton and a relatively fast (and large) soliton
($\lambda =0.21 <2/9$),
whereas on the right panel, we combine a pair of equal-width
flatons and follow their interaction with a soliton. Note that
by the time the soliton collides with the flaton, the latter
has already settled into its ultimate form.

The soliton-flaton interactions clearly
indicate that flatons are more than merely approximate solutions and
there should be an underlying analytical structure.
Indeed, it was always there, it just had to be unfolded. The 'trick' is not
to take the $\lambda \rightarrow 2/9$ limit but to focus on
a narrow layer squeezed between
the upper solitons range, Eq. (\ref{TW2modkdv}), and their barrier where
 kink/antikink form. Flatons saga can be read from the left panel of
Fig. 4 where we draw
profiles of $u$; whereas at small velocity $\lambda$, solitons
have their usual peaked shape,
upon approaching their upper speed limit, their shape changes
drastically and rather
than to grow with $\lambda$, a slightest increase in
velocity causes them to widen very quickly.

\begin{figure}
\label{comb89}
\noindent \begin{centering}
\hspace*{-4cm}
\vspace*{-4cm}
\includegraphics[width=55pc]{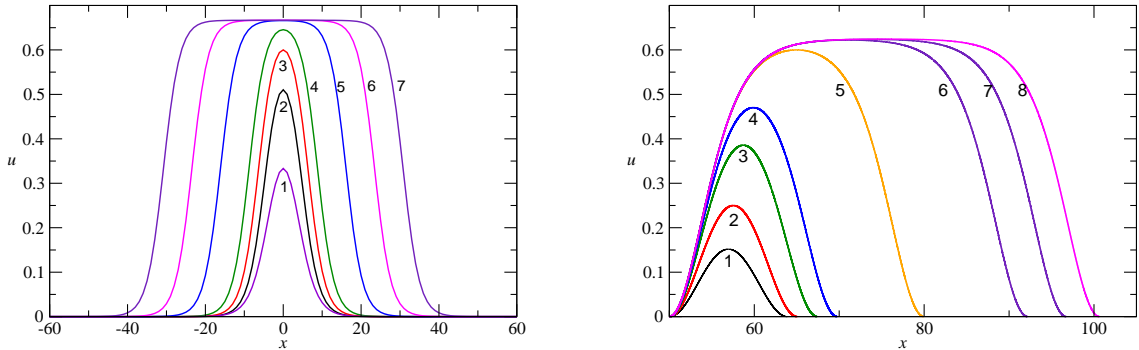}
\par\end{centering}
\vspace*{-21cm}
\caption{
Left panel: Soliton's profile as an ascending function of its velocity $\lambda=1/6,~0.21,~0.22,~0.222,~2(1-10^{-6})/9,~2(1-10^{-9})/9$ and
$2(1-10^{-12})/9$,
respectively. Note the structural change that soliton's shape
undergoes as $\lambda\rightarrow 2/9$; their top flattens,
the amplitude hardly changes and they
turn into a sequence of ever widening flatons.
Right panel:
Profiles of G(2) compactons, Eq. (\ref{23n}) with $n=2$,
as a function of $\lambda$. When $\lambda \rightarrow 15/64$, compactons turn into flatons.
Cases 1 to 8 correspond to $\lambda=0.1,~0.15,~0.2,~0.22,~0.234,~0.23437,
~0.234374$, and $0.23437475$, respectively.
}
\end{figure}

To 'extract' those features from solitons formula (\ref{TW2modkdv}),
we define
$$ \lambda_{f} = \frac{2}{9}\Big(1-\eps^{2}\Big)~~~~
{\rm{where}}~~ 0<\epsilon \ll 1,$$
then, in terms of $\epsilon$, Eq. (\ref{TW2modkdv}) reads
\begin{equation}
u=\frac{2}{3} \left( \frac{1-\eps^{2}}{1+\eps\cosh{\left[\sqrt{\lambda_f}
(x-  \lambda_{f}t)\right]}}\right)
\label{ueps}
\end{equation}
and thus $u_{max} =\frac{2}{3}(1-\eps)$.

Comparing the exact solution with its approximation (\ref{appflaton}) we find
 their difference to be ${\cal{O}}\left(\epsilon^{2}\right)$ and the extent of soliton's profile widening and flattening
may be expressed via $x_{1/2}$ where the soliton's amplitude
has decreased by half, i.e., when
\begin{equation}
x_{1/2} \simeq \frac{3}{\sqrt{2}}\ln{\frac{2}{\eps}}.
\label{x0eps}
\end{equation}

Thus, velocity change $\sim 1-\epsilon^{2}$  and amplitude change
$\sim 1-\epsilon$, cause the corresponding flaton to widen as $\sim \ln{1/\eps}$. \\

That 1D flatons went for so long unnoticed may perhaps be due to
their very narrow domain of attraction.
The left panel of Fig. 5 displays emergence of a flaton out
of a {\it carefully orchestrated initial excitation}.
With one exception we have found this feature to be typical of
all studied cases.
Therefore, it would have been misleading to start unfolding
flatons with the left panel of Fig. 5 which is
an end result of numerous numerical experiments and
was 'extracted' after the mere existence of
flatons was firmly established.\\

{\bf Compact Flatons.} Turning to flatons with a compact support we
start with
\\
(i) \leftline{{\bf The G(n=2) equation.}}
Without the defocusing cubic term,
Eq. (\ref{23n}) reduces to the K(2,2) equation \cite{RosHym}
with the underlying compacton
\begin{equation}
u=\frac{4\lambda}{3}\cos^{2}\left(\frac{s}{4}\right)H(2\pi-|s|)~~~
{\rm {where}}~~ s=x-\lambda t
\label{CompK22}
\end{equation}
($H(y)$ is the Heaviside function),
being its basic solitary mode with a compact support. Though K(2,2) is not
integrable in the conventional sense, its interactions are
remarkably clean \cite{RosHym,RZ}.

To derive the compact solitary waves let
$u=u(s)$ where $s=x-\lambda t$ and integrate
Eq. (\ref{23n}) with $n=2$ twice to obtain
\begin{figure}
\label{comp232}
\noindent \begin{centering}
\hspace*{-4cm}
\vspace*{-4cm}
\includegraphics[width=55pc]{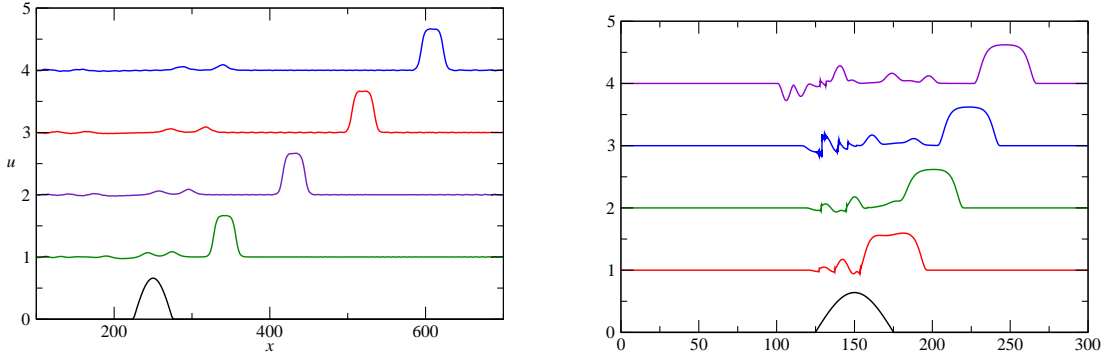}
\par\end{centering}
\vspace*{-21cm}
\caption{
Evolution of carefully tailored initial excitations. Left panel: G(1).
$u_0(x)= 0.66 \cos{(x/16)}H(8\pi-|x|)$
which begets as its main response the G(1) flaton, see Eq. (\ref{ueps});
the displayed times are $t=0, 400, 800, 1200$ and $t=1600$.
Right panel: G(2). $u_0=0.66 \cos{(x/16)}H(8\pi-|x|)$
which begets as its main response the G(2) compact flaton.
The  displayed interactions are at time intervals of $\Delta t=100$.}.
\end{figure}

\begin{equation}
 u^{2}\Big(u^{2}_{s}+P(u;\lambda)\Big)=0~~~{\rm {with}}~~~
P(u;\lambda)=-\frac{\lambda}{3}u+\frac{1}{4}u^{2}-\frac{1}{5}u^{3}
\end{equation}
where $u^{2}$ is kept out of the bracket to stress the singular nature
at $u=0$ which induces compactness. In terms of $u=\frac{4\lambda}{3}W^{2}$,
 we obtain
 \begin{equation}
4 W^{2}_{s} =1-W^{2}+\gamma W^{4} = \gamma(W^{2}_{-}-W^{2})(W^{2}_{+}-W^{2})
\label{elfun}
 \end{equation}
where
$$
~~\gamma=\frac{16}{15}\lambda~~~
{\rm {and}}~~W^{2}_{\pm}=\frac{2}{1\mp\sqrt{1-4\gamma}}~~.
$$

Insofar as $\gamma<1/4$ or $\lambda <15/64$, we obtain compactons.
Their profiles displayed in the right panel of Fig. 4
may be expressed via elliptic functions, but were determined
numerically solving the first order ode Eq. (\ref{elfun}) via evaluation
of the primitive of the inverse potential function $P(u; \lambda)$.
Note that unlike the K(2,2), the width of G(2) compactons
depends on their velocity.
When $\gamma=1/4$, $W_{-}=W_{+}$, compactons dissolve and
semi-compact kink  and anti-kink vanishing at $u=0$ emerge
 \begin{equation}
 u=\frac{5}{8}\tanh^{2}\left(\frac{\pm s}{4\sqrt{2}}\right)
H(\pm s),~~~s=x-\frac{15}{64}t.
\label{tanh}
  \end{equation}

The right panel of Fig. 5 illustrates the emergence of a compact flaton
out of a carefully tailored initial excitation and, as in G(1)
shown in the left panel of Fig. 5,
is the end result of numerous numerical experiments.
The left panel of Fig. 6 displays interaction of the G(2) compact flaton
with a G(2) compacton. The flaton emerges unchanged, but the compacton sheds
some of its mass.\\

\leftline {\bf (ii) The G(n=3) equation.}

For the present case we have an exact solution \cite{RO}
\begin{equation}
u= \frac{3}{5}\left[1-\sqrt{1-\frac{25\lambda}{6}}\cosh{(\frac{s}{3})}\right]_{+}
\label{G3comp}
\end{equation}
representing  for $\displaystyle 0 < \lambda < 6/25$ a compacton.
As elsewhere, flatons emerge near the edge of their upper range
$$ \lambda_{f}=\lambda_{max}(1-\epsilon^{2})~~~{\rm {where}}
~~\lambda_{max}=\frac{6}{25}, $$
yielding
\begin{equation}
u= \frac{3}{5}\left[1-\epsilon\cosh{\left(\frac{x-\lambda_{f}t}{3}\right)}\right]_{+}.
\end{equation}

\begin{figure}
\label{flatcompG2}
\noindent \begin{centering}
\hspace*{-4cm}
\vspace*{-4cm}
\includegraphics[width=55pc]{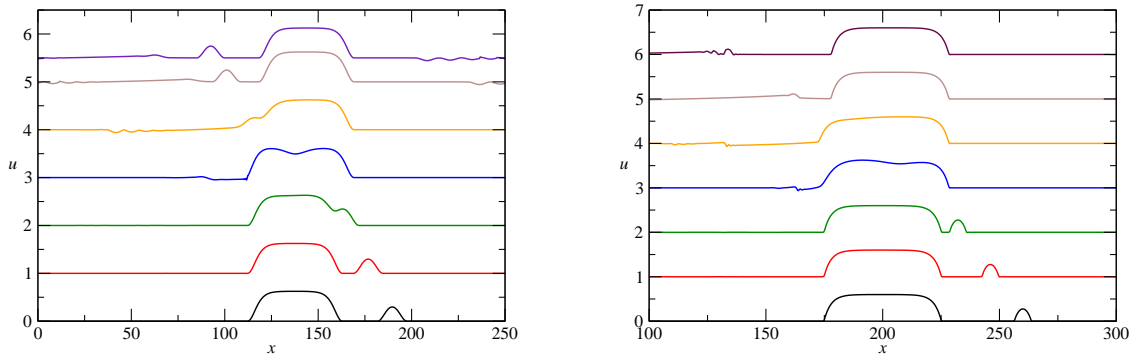}
\par\end{centering}
\vspace* {-21cm}
\caption{
Left panel: G(2)- Interaction of a $\lambda=0.1$ compacton with
a $\lambda=0.23437475$ compact flaton.
Right panel:
G(3)- Interaction of a $\lambda=0.17$ compacton with a
$\lambda= 0.23999995$ compact flaton.
In both cases, flatons reemerge from the encounter unchanged,
but in both cases the smaller
compacton sheds some of its mass. The snapshots in both panels
correspond to time intervals of $200$.
}
\end{figure}

The right panel of Fig. 6 displays G(3) compact flaton
colliding with a G(3) compacton. The flaton emerges unaffected,
but the compacton loses a big chunk of its mass.
We note that unlike all other studied cases, G(3)
flatons emerge quite 'naturally' out of generic initial excitations (not shown).
This could be attributed, at least in part, to the fact that since for G(3) flatons $u<0.6$, the convection velocity $C(u)$ remains always positive.
We shall further comment on this issue in the last section.


\section*{3. Multi-dimensional Flatons}

We now turn to the
multi-dimensional extension of the Gardner equation (\ref{ND})
and seek spherically symmetric solitary waves $U(r=\sqrt{s^{2}+y^{2}+ z^{2}})$,
 $s=x-\lambda t$, that propagate in x-direction.
{ Hereafter we address only the $n=1$ case. Other cases are deferred to a future publication.}

After one integration
 \begin{equation}
  -\lambda U(r) + {c_+ U^{\kappa}-c_- U^{\ell}+ \frac{d-1}{r}U^\prime(r)+
U^{\prime \prime}} = 0, ~~~\kappa, \ell >1.
\label{ode}
\end{equation}
Assuming $r$ to play the role of time, $(d-1)/r$ may be looked upon
as a time-dependent 'friction coefficient'
which decays with time. Another formal integration yields
\begin{equation}
\frac{1}{2}U'^{2} + P(U) + (d-1)\int{\frac{dr}{r}{U'^{2}}}=E_{0}~~~
\end{equation}
{\rm {where}}
\begin{equation}
P(U)= -\frac{\lambda}{2}U^{2} + { c_+\frac{U^{1+\kappa}}{1+\kappa} - c_-} \frac{U^{1+\ell}}{1+\ell}
\label{Pu}
\end{equation}
with $E_{0}$ playing the role of an effective total energy.

Consider again the potential landscape on Fig. 1.
To overcome the 'friction' present whenever $d>1$,
the starting point $u_{0}$ from which the "particle"
starts its descent has to be moved up the potential hill.
As in 1D case, if potential's peak is close enough to the u-axis then
the spherical solitary wave will be a flaton, but now things are a bit
more involved for if the particle does not have initially
a sufficient potential energy,
the friction may stop it prior to its arrival to the origin. This
difficulty is resolved noting that since on potential's top
the particle can rest indefinitely, thus,
 being close enough to the top, enables the particle to delay
its descent until the time-dependent friction $(d-1)/r$ becomes
sufficiently suppressed, so that it could not prevent the particle
from arriving to the origin. The longer the particle 'waits',
the longer becomes its flat top.

\begin{figure}
\label{fig231}
\noindent \begin{centering}
\hspace*{-2cm}
\includegraphics[width=50pc]{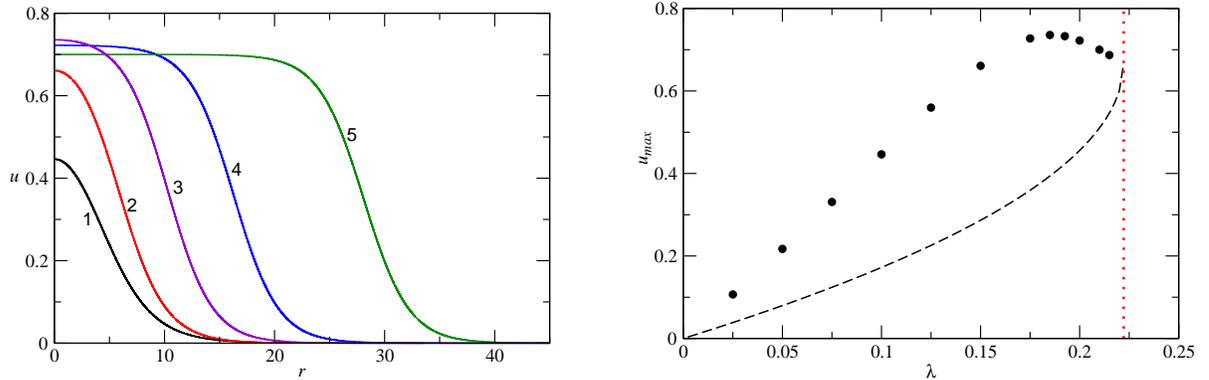}
\par\end{centering}
\vspace* {-23cm}
\caption{
$G_{3}^{1}(2,3)$, Eq. (\ref{ND}).
Left panel: Displays 1-5 correspond to $\lambda=0.1,~0.15,~0.185,~0.2$
and $0.21$, respectively, with
the latter two being flatons, whereas the corresponding 1D patterns in
Fig. 4 are still very much conventional solitons.
Note the non-monotone dependence of solitons
amplitudes on their speed, explicitly shown by the circles
on the right panel,
which happens concurrently with the emergence of flatons;
for though soliton's peak ascends with $\lambda$ toward potential's top,
this does not compensate for potential's descent toward the u-axis.
The dashed curve displays the $\lambda$-dependence of the 
maximal amplitude, $u_{max}$,
of the 1D solitons/flatons, see Eq. (\ref{umax}).
The vertical dotted line at $\lambda=2/9$ marks the upper limit
of the admissible propagation speeds.
}
\end{figure}

Left panel of Fig. 7 is the 3D counterpart of Fig.'s 4 left panel
and, as expected, shows that 3D flatons exist for a much wider range of
velocities. The right panel of Fig. 7 displays the non-monotone
{dependence}
of solitons peak $u_{max}(\lambda)$ on its velocity.
Insofar that solitons peak is far from potential's top then, as in 1D,
$u_{max}(\lambda)$ increases with $\lambda$. However, since concurrently with  $\lambda$ increase, potential's
top decreases, from a certain velocity on solitons
peak comes close enough to potential top so that flatons form.
From now on, though with a further increase in velocity, soliton's
peak will be even closer to the top, this will not compensate for
potential's descent toward the $u$-axis, with the net effect
that flatons amplitude, as clearly seen in Fig. 7 decreases.
As velocity approaches its upper bound, $u_{max}(\lambda)$ approaches
from above the amplitude (=2/3) of 1D kinks.

\begin{figure}
\label{potenQKDV}
\noindent \begin{centering}
\vspace*{4mm}
\includegraphics[width=17pc]{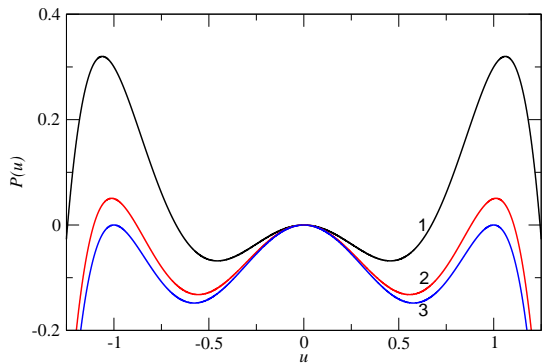}
\par\end{centering}
\vspace* {-3.5mm}
\caption{
The $G_{1}^{1}(3,5)$ potential, Eq. (\ref{P351}):
$P=-\frac{u^{2}}{2}\big(\lambda -2u^{2} +u^{4}\big)$.
Cases 1-3 correspond to  $\lambda=0.7$, $0.95$ and $1$,
respectively. The second case begets flatons. The third - kinks/antikinks.
}
\end{figure}

\begin{figure}
\label{d5d6}
\noindent \begin{centering}
\hspace*{-4cm}
\vspace*{-4cm}
\includegraphics[width=55pc]{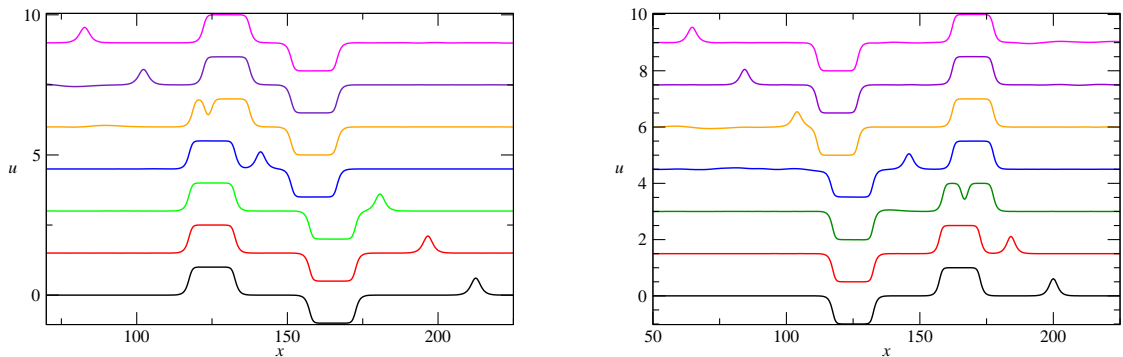}
\par\end{centering}
\vspace* {-22cm}
\caption{
$G_{1}^{1}(3,5)$, Eq. (\ref{mod35kdv}). Left panel:
Interaction of a $\lambda=0.6$ soliton, Eq. (\ref{351}), with
flaton-antiflaton formation, $\epsilon=10^{-6}$, Eq.(\ref{351flaton}).
Right panel: The antiflaton and the
flaton have switched their positions causing an opposite post
collision dislocation.
Both displays are in {a} reference frame moving
with $\lambda_f= 1-\epsilon^2$. 
On both panels the shown displays, from the bottom up, 
are at time intervals of $\Delta t=40$.
}
\end{figure}


{\bf The} $\mathbf{G_{3}^{1}(3,5)}$. 
It is convenient to address first its 1D variant
\begin{equation}
G_{1}^{1}(3,5):~~~u_t + \left(4u^3-3u^5\right)_x+u_{xxx}=0, \label{mod35kdv}
\end{equation}
with its underlying traveling waves potential, see Fig. 8,
\begin{equation}
 P(u)=-\frac{u^{2}}{2}\big(\lambda -2u^{2} +u^{4}\big)
\label{P351}
\end{equation}
which for $0 < \lambda < \lambda_{max} =1$  admits solitary solutions
\begin{equation}
u=\pm \frac{\sqrt {\lambda}}{\sqrt{1+
\sqrt{1-\lambda}\cosh\left[2\sqrt{\lambda}(x-\lambda t)\right]}},
\label{351}
\end{equation}
peaking at
\begin{equation}
u_{max}^{2} = 1-\sqrt{1-\lambda},
\end{equation}
and kinks when $\lambda=1$
\begin{equation}
u=\pm \frac{1}{\sqrt{1+\exp{\left(2(x-t)\right)}}}.
\end{equation}

Let $\lambda_{f}= (1-\epsilon^{2}) \lambda_{max}
=1-\epsilon^{2}$, $0<\epsilon<<1$, then flatons follow
\begin{equation}
u=\pm \frac{{ \sqrt {1-\epsilon^2}}}{\sqrt{1+\epsilon
\cosh{\left[2\sqrt{1-\epsilon^{2}}(x-\lambda_{f} t)\right]}}}.
\label{351flaton}
\end{equation}
Here, the convection velocity $C(u)= 12u^{2}-15u^{3}$ reverses its direction
 and whenever $\sqrt{0.8}< u $, it acts in a
direction opposite to flatons motion.
Yet, as can be seen in Fig. 9, this does not seem to have a direct
impact on their dynamics: both flatons and anti-flatons seem very robust,
and their interaction with solitons is fairly clean, though
 solitons seem to lose some mass. Also, since the collision causes
the flaton (anti-flaton) to move to the right (left),
the distance between the flaton and anti-flaton decreases when
the anti-flaton is hit first, see the left panel, but
increases when the flaton is the one to be hit first, see the right panel. However,  convection's reverse of direction within flatons amplitude range manifests itself otherwise;
even with a careful tailoring of initial excitations we did not observe flatons emerge. They had to be 'planted' ab initio.

\begin{figure}
\label{fig351}
\noindent \begin{centering}
\vspace*{4mm}
\hspace*{-1.5cm}
\includegraphics[width=45pc]{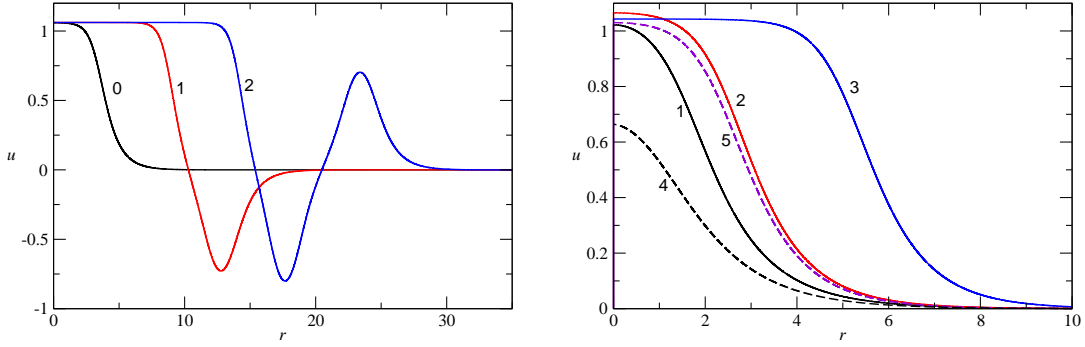}
\par\end{centering}
\vspace* {-21cm}
\caption{
$G_{3}^{1}(3,5)$. Left panel: Display of the first three $\lambda=0.7$ 3D
flatons. Right panel: $\lambda=0.4,~0.6$
and $0.8$ 3D solitons marked by 1, 2 and 3, respectively and
$\lambda=0.8$ solitary waves in one and two dimensions marked by 4 and
5, respectively.
}\end{figure}

A far richer scenario awaits us in the spherical
extension of the problem. Let $u=u_{0}(\lambda)$ be the initial
position of a "particle" on potential's particular positive branch
descending from there to $u=0$. Now, since the potential is
symmetric, a further climb on the positive hill will
lead to point $u_{1}$ where the descending particle has a
sufficient potential energy to overcome the friction,
pass through the origin, make one round in the
negative well and settle at the origin. A further climb along the potential
leads to point $u_{2}$ which equips the "particle" with adequate
potential energy to make two rounds
between the two potential wells prior to its settling at the origin.

For an unbounded potential like $P(u)=-u^{2}+u^{4}$, the above procedure would
beget an infinite sequence of
$N$-nodal solitons, however, since the relevant potential
$P(u)$ with $\kappa=2$ and $\ell=4$ peaks at
\begin{equation}
u^{2}_{P_{max}}=\frac{2}{3}\Big(1+\sqrt{1-\frac{3\lambda}{4}}\Big),~~\lambda\leq 1,
\end{equation}
a climb along potential's specific branch
can  proceed only so far. Yet, in this case as well there is a
sequence of multi-nodal solitons, but now they condense
near the potential's top.
Again, a "particle" close enough to the top may wait there
until 'friction' becomes sufficiently suppressed and let the particle execute
exactly $N$-oscillations
prior to its settling at the origin. One thus derives
a sequence of modes condensing near
potential's top with $(N+1)$-nodal soliton "waiting" for a
longer 'time' and thus acquiring a longer quasi-plateau
than its $N$-nodal predecessor prior to its descent.
The left panel of Fig. 10
displays a $\lambda=0.7$ example of first three such
modes.  Note that the difference between the initial values of
$u_{0}$, $u_{1}$ and $u_{2}$ is smaller than $10^{-8}$.
Also, since  the 3D 'friction' is larger than the planar one,
the spherical particle has to hover near the top for a longer 'time' which
results in both slightly higher initial amplitude than the
respective 2D case, and a longer quasi-plateau, see the right plate of
Fig. 10. { The latter also shows that with an increase in $\lambda$
solitons transform to flatons emerging already at
$\lambda \approx 0.6 \lambda_{max}$.}

\begin{figure}
\label{poten3}
\noindent \begin{centering}
\vspace*{4mm}
\includegraphics[width=20pc]{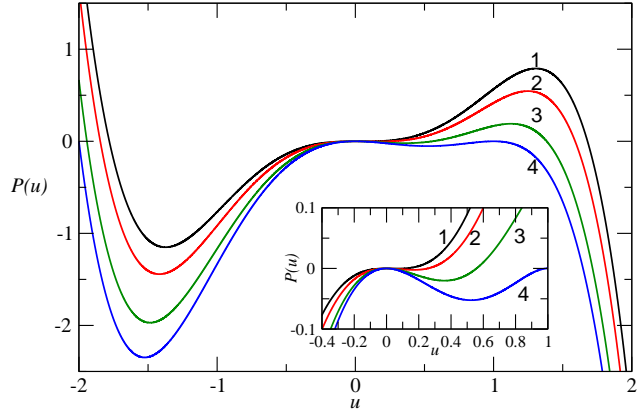}
\par\end{centering}
\vspace* {-3.5mm}
\caption{
The potential $P= -\frac{\lambda}{2} u^{2} + u^{3} - {\frac13} u^{5}$
of Eq. (\ref{ND2}) {with
$\kappa=2$, $\ell=4$, $c_{+}=3$ and $c_{-}=5/3$},
for $\lambda=0.2, ~0.5, ~1$ and $4/3$  marked by 1 - 4,
respectively. The inset: vicinity of the origin.
}
\end{figure}

\begin{figure}
\label{241L02}
\noindent \begin{centering}
\vspace*{4mm}
\includegraphics[width=40pc]{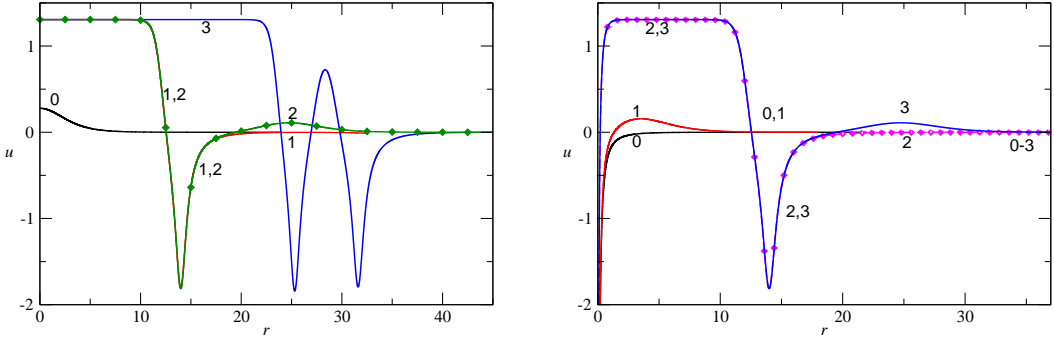}
\par\end{centering}
\vspace* {-18cm}
\caption{
$\lambda=0.2$. Bright and dark spherical solitons of Eq. (\ref{ND2}).
Left panel: The first four bright modes. Note that up to $r \approx 20$,
modes 1 and 2 coincide visually.
Right panel: The first four dark modes. Notably, apart of a very narrow
layer near dark modes origin where { $u_{0}= -6.54976 \ldots$,
$u_{1}=-7.02246$ \ldots,
$u_{2}=-15.821105239848875 \ldots $ and $u_{3}=u_2-5.21 \times 10^{-13}$,}
bright and dark modes look very similar!
}
\end{figure}

Thus, whereas in 1D, for flatons to emerge we need the potential
{ top} to come
close enough to the $u$-axis, in higher dimensions there is a joint action: on one hand one has to climb toward
potential's top on the other, the top comes down which results in flatons forming in a much wider
velocities range, c.f., the G(1) case,
whereas in a symmetric potential it always assures flatons, though
not necessarily starting with the basic $N=0$ mode. 
Hereafter we shall refer to either a soliton or flaton as an
$N$-mode if it crosses $u=0$, $N$ times.

The last case to be considered amalgamates the
two previous cases. Let $\kappa=2$ and $\ell=4$.
Without loss of generality we assume $c_{+}=3$ and $c_{-}=5/3$. Thus
\begin{equation}
 G_{3}^{1}(2,4):~~~u_{t} + \left(3u^{2}-\frac53 u^{4}\right)_{x}+(\nabla^{2}u)_{x}=0. \label{ND2}
\end{equation}
The underlying potential of its travelling waves, see Fig. 11,
\begin{equation}
P(u)= -\frac{\lambda}{2} u^{2} + u^{3} - {\frac13} u^{5},
\end{equation}
has two very asymmetric wells. Its positively valued branch
admits the so-called bright solitons, i.e., solitons which assume a
positive value at the origin, in a bounded velocity range,
whereas the negative branch, $u<0$, supports dark ones,
i.e., solitons which assume a negative
value at the origin, for any $\lambda$. In more detail:

1) Bright modes.
Since potential's negative branch reflects all particles coming from the positive side,
they may then condense near potential's top.
Left panel of Fig. 12 displays an example of the first{ four}
spherical modes from the basic $N=0$ till $N=3$.

2) Dark modes: Let $u=u_{0}<0$ denote the amplitude of its
0-nodal dark soliton. Now let $u=u_{top}$ be a point on the negative branch such
that a "particle" released from $P(u_{top})$  lands on the top of
potential's positive branch. Clearly, any initial position
$u<u_{top}$ will cross potential's top and roll to infinity.
Thus, the dark modes range is limited to the
$u_{top}<u<u_{0}<0$ strip. One may now attempt to repeat the condensation
scenario of the previous case near the positive top by
launching particles sufficiently
close to $u=u_{top}$, expecting them to land near the
top and wait there until the 'friction' is subdued and
then repeat the spiel: descend to the valley and execute $N$ oscillations prior
to settling at the origin. Such solution would start as a dark entity
followed by a {\it flat part} and conclude with an oscillatory tail,
c.f., right plate of Fig. 12 where four such modes are displayed.

Finally, we pause to note that all multi-dimensional solutions
presented in this section were numerically found solving
Eq. (\ref{ode})
amended with boundary
conditions $U^\prime(r=0)=0$, and $ U(r \to \infty) \to 0$
for the relevant values of $\kappa$, $\ell$ and $c_{\pm}$
via a shooting method based on fourth-order Runge- Kutta 
scheme with a typical $\Delta r= O\left(10^{-5}\right)$ step size.
In spite of using quadruple-precision, 128 bit computations,
we were unable to unfold numerically more than three sign-changing modes,
with the difficulty escalating very quickly as $\lambda$ increases
toward the limiting value beyond which the solution diverges.
To appreciate the computational difficulty, note
that the convergence to the far-field condition $r \to \infty$ requires
exceedingly small changes in $u(r=0)$, for instance at the $17^{th}$ digit
in the case of the $0^{th}$-mode for $\lambda=0.215$ shown
in the right panel of Fig. 7. Part of the difficulty stems from the fact that
unlike the previous cases which were
initiated from the positive top's vicinity, and we were able to 
control their starting
point, and thus the waiting time, with the starting point located
on the negative branch, we have no
direct control over the landing point on the positive branch,
and apart of the first four cases, searching for 
higher modes we always either ended in the negative 
well or escaped all together the potential hill.
Though one cannot preclude that higher $N$ modes could be
unfolded with a much higher precision, a truly-high $N$
modes seem to be out of reach for both dark and bright modes.

\section*{4. Summary}

The present work focuses on unfolding flatons - flat-top -
solitons which emerge in dispersive systems whenever
solitons range is bounded from above.
Whereas in two or three dimensions any solution hovering long enough close to potential's peak, or better yet. emerging there,
may be 'colonized' by a flaton(s), for 1D flatons to emerge,
potential's peak has to be brought down sufficiently close to the u-axis,
which takes place only when solitons speed approaches the top of its range.
This makes 1D flatons far more delicate affair;
they are very sensitive to even a minute change of their velocity which
causes a dramatic change of their width.

Finally, we comment upon 1D flatons domain of attraction
which with one exception is very narrow, as one can witness
from the carefully orchestrated initial excitations in Fig. 5
which beget flatons and were done posteriori after flatons
existence was already established. { Though, as we have stressed,
the existence of flatons} bears no consequence as to their
robustness once they emerge or are planted ab initio.
Two cases are to be contrasted: whereas for $G_{1}^{1}(3,5)$
we were unable to see flatons emerging from any reasonable
initial conditions, for G(3)
they emerge quite naturally from any generic initial excitation.
The only plausible explanation that comes to mind is
 that whereas G(3) flatons amplitude evolves within cooperative
domain of convection, in the $G_{1}^{1}(3,5)$ case flatons
amplitude, $0\leq u<1$, falls within a regime where
convection reverses its direction, $C(u=\sqrt{0.8})=0$, and opposes
in part motion of the flaton. G(1) is a borderline case:
convection vanishes at $u=2/3$ which is flatons' upper bound,
whereas G(2) represents an intermediate case
between G(1) and G(3).
This perhaps answers in part why unlike the spherical 
flatons which were already noted before in the complex
field theory \cite{Q-balls, Q-balls2}, 1D flatons have 
hitherto escaped our attention. In closing, we note 
that flatons in spatially discrete system will be addressed in \cite{RoPi}.

\section*{Declaration of Competing Interest}

The authors declare that they have no known competing financial
interests or personal relationships that could have appeared to
influence the work reported in this paper.

\section*{Acknowledgments}

A. O. was supported in part by the David T. Siegel Chair in Fluid Mechanics
and by the ISF Grant no. 356/18.

\newpage

\section*{Figures captions}

Fig.1:
Display of the potential
$P_1(u)=-\frac{1}{2}\lambda u^{2}+\frac{1}{3}u^{3}-\frac{1}{4}u^{4}$ for
Eq. (\ref{G1}) with $\lambda=1/9$, $7/36$ and $2/9$, cases 1-3, respectively.
Flatons emerge in the domain between cases 2 and 3.
In case 3 wherein $P=P'=0$, there are no solitons. Instead
kink and/or anti-kink emerge.

Fig. 2:
Comparison between the
approximate shape (solid lines) of the G(1) flaton, Eq. (\ref{appflaton}),
and the exact solution (dashed lines) for a several values of
{$x_0$ (or $\epsilon$, see Eq. (\ref{x0eps}))}:
1- $x_0=8.0 $ ($\Rightarrow \epsilon =0.05$),
2- $x_0=11.27$ ($\Rightarrow  \epsilon =10^{-2}$),
3- $x_0=16.13$ ($\Rightarrow \epsilon =10^{-3}$),
4- $x_0=21.01$ ($\Rightarrow \epsilon =10^{-4}$).
Notably, apart of the first pair with visible difference at the top
between the exact solution and its approximation,
in other cases, they are truly indistinguishable.

Fig. 3:
Interaction between G(1) flatons and a soliton.
Left panel: Space-time display of the interaction of a wide
flaton, $x_0=25$, with a  $\lambda=0.21$ soliton,
shown here and elsewhere in a reference frame moving with the kink velocity of the flaton.
From the bottom to the top; the
time ranges from $t=0$ to $t=10^{4}$ with time intervals of  $1000$.
Right panel:
Space-time display of  the interaction between two
equal flatons, $x_0=20$, with a $\lambda=0.21$ soliton.
From the bottom to the top, the time ranges from $t=0$
to $12000$ with time intervals of $1000$ and finally $t=16000$.

Fig. 4:
Left panel: Soliton's profile as an ascending function of its velocity $\lambda=1/6,~0.21,~0.22,~0.222,~2(1-10^{-6})/9,~2(1-10^{-9})/9$ and
$2(1-10^{-12})/9$,
respectively. Note the structural change that soliton's shape
undergoes as $\lambda\rightarrow 2/9$; their top flattens,
the amplitude hardly changes and they
turn into a sequence of ever widening flatons.
Right panel:
Profiles of G(2) compactons, Eq. (\ref{23n}) with $n=2$,
as a function of $\lambda$. When $\lambda \rightarrow 15/64$, compactons turn into flatons.
Cases 1 to 8 correspond to $\lambda=0.1,~0.15,~0.2,~0.22,~0.234,~0.23437,
~0.234374$, and $0.23437475$, respectively.

Fig. 5:
Evolution of carefully tailored initial excitations. Left panel: G(1).
$u_0(x)= 0.66 \cos{(x/16)}H(8\pi-|x|)$
which begets as its main response the G(1) flaton, see Eq. (\ref{ueps});
the displayed times are $t=0, 400, 800, 1200$ and $t=1600$.
Right panel: G(2). $u_0=0.66 \cos{(x/16)}H(8\pi-|x|)$
which begets as its main response the G(2) compact flaton.
The  displayed interactions are at time intervals of $\Delta t=100$.

Fig. 6:
Left panel: G(2)- Interaction of a $\lambda=0.1$ compacton with
a $\lambda=0.23437475$ compact flaton.
Right panel:
G(3)- Interaction of a $\lambda=0.17$ compacton with a
$\lambda= 0.23999995$ compact flaton.
In both cases, flatons reemerge from the encounter unchanged,
but in both cases the smaller
compacton sheds some of its mass. The snapshots in both panels
correspond to time intervals of $200$.

Fig. 7:
$G_{3}^{1}(2,3)$, Eq. (\ref{ND}).
Left panel: Displays 1-5 correspond to $\lambda=0.1,~0.15,~0.185,~0.2$
and $0.21$, respectively, with
the latter two being flatons, whereas the corresponding 1D patterns in
Fig. 4 are still very much conventional solitons.
Note the non-monotone dependence of solitons
amplitudes on their speed, explicitly shown by the circles
on the right panel,
which happens concurrently with the emergence of flatons;
for though soliton's peak ascends with $\lambda$ toward potential's top,
this does not compensate for potential's descent toward the u-axis.
The dashed curve displays the $\lambda$-dependence of the
maximal amplitude, $u_{max}$,
of the 1D solitons/flatons, see Eq. (\ref{umax}).
The vertical dotted line at $\lambda=2/9$ marks the upper limit
of the admissible propagation speeds.

Fig. 8:
The $G_{1}^{1}(3,5)$ potential, Eq. (\ref{P351}):
$P=-\frac{u^{2}}{2}\big(\lambda -2u^{2} +u^{4}\big)$.
Cases 1-3 correspond to  $\lambda=0.7$, $0.95$ and $1$,
respectively. The second case begets flatons. The third - kinks/antikinks.

Fig. 9:
$G_{1}^{1}(3,5)$, Eq. (\ref{mod35kdv}). Left panel:
Interaction of a $\lambda=0.6$ soliton, Eq. (\ref{351}), with
flaton-antiflaton formation, $\epsilon=10^{-6}$, Eq.(\ref{351flaton}).
Right panel: The antiflaton and the
flaton have switched their positions causing an opposite post
collision dislocation.
Both displays are in {a} reference frame moving
with $\lambda_f= 1-\epsilon^2$.
On both panels the shown displays, from the bottom up,
are at time intervals of $\Delta t=40$.

Fig. 10:
$G_{3}^{1}(3,5)$. Left panel: Display of the first three $\lambda=0.7$ 3D
flatons. Right panel: $\lambda=0.4,~0.6$
and $0.8$ 3D solitons marked by 1, 2 and 3, respectively and
$\lambda=0.8$ solitary waves in one and two dimensions marked by 4 and
5, respectively.

Fig. 11:
The potential $P= -\frac{\lambda}{2} u^{2} + u^{3} - {\frac13} u^{5}$
of Eq. (\ref{ND2}) {with
$\kappa=2$, $\ell=4$, $c_{+}=3$ and $c_{-}=5/3$},
for $\lambda=0.2, ~0.5, ~1$ and $4/3$  marked by 1 - 4,
respectively. The inset: vicinity of the origin.

Fig. 12:
$\lambda=0.2$. Bright and dark spherical solitons of Eq. (\ref{ND2}).
Left panel: The first four bright modes. Note that up to $r \approx 20$,
modes 1 and 2 coincide visually.
Right panel: The first four dark modes. Notably, apart of a very narrow
layer near dark modes origin where { $u_{0}= -6.54976 \ldots$,
$u_{1}=-7.02246$ \ldots,
$u_{2}=-15.821105239848875 \ldots $ and $u_{3}=u_2-5.21 \times 10^{-13}$,}
bright and dark modes look very similar!

\end{document}